\documentclass[prl,twocolumn,amsmath,amssymb,showpacs]{revtex4} %PRL format

\usepackage{epsfig}
\usepackage[english]{babel}
\usepackage[normalem]{ulem} %emphasize weiterhin kursiv

\newcommand{\beq}{\begin{equation}}
\newcommand{\eeq}{\end{equation}}
\newcommand{\beqarr}{\begin{eqnarray}}
\newcommand{\eeqarr}{\end{eqnarray}}

\begin{document}

\title{Zipf law in the popularity distribution of chess openings}

\author{Bernd Blasius$^{1}$}
\author{Ralf T{\"o}njes$^{2,3}$}
\affiliation{${}^{1}$ ICBM, University Oldenburg, 26111 Oldenburg, Germany}
\affiliation{${}^{2}$ Institut f\"ur Physik, Universit\"at Potsdam, 14415 Potsdam, Germany}
\affiliation{${}^{3}$ Ochadai Academic Production, Ochanomizu University, Tokyo 112-8610, Japan}

\begin{abstract}  % no more than 600 characters
We perform a quantitative analysis of extensive chess databases and show that 
%there are simple statistical laws underlying the choice of opening moves.
the frequencies of opening moves are distributed
according to a power-law with an exponent that increases linearly with the game depth, whereas
the pooled distribution of all opening weights follows Zipf's law with universal exponent.
We propose a simple stochastic process that is able to capture the observed playing
statistics and show that
the Zipf law arises from the self-similar nature of the game tree of chess. 
Thus, in the case of hierarchical
fragmentation the scaling is truly universal and independent of a particular generating
mechanism. Our findings are of relevance in general processes with composite decisions.
%where beyond a critical number of decision there is a transition to inequal distributions.
\end{abstract}

\maketitle
Decision making refers to situations where individuals have to select a course of action among multiple alternatives \cite{Simon47}. 
Such processes are ubiquitous, ranging from one's personal life 
to business, management and politics  and take a large part in shaping our life and society. 
Decision making is an immensely complex process and, given the number of factors that influence each choice,  a quantitative understanding 
in terms of statistical laws remains a difficult and often elusive goal.
Investigations are complicated by the shortage of reliable data sets, since information about human behavior is often difficult to be quantified and not easily available in large numbers, whereas decision processes typically involve a huge space of possible courses of action.  Board games, such as chess, provide a well-documented case where the players in turn select their next move among a set of possible game continuations that are determined by the rules of the game. 

Human fascination with the game of chess is long-standing and pervasive
\cite{Murray02}, not least due to the sheer infinite richness of the game. 
The total number of different games that can be played, i.e., 
the game tree complexity of chess, has roughly been estimated as 
the average number of legal moves in a chess position 
to the power of the  length of a typical game,
yielding the Shannon number $30^{80} \approx 10^{120}$ \cite{Shannon50}.  
Obviously only a small fraction of all possible games can be realized in actual play. But even during the first moves of a game, when the game complexity is still manageable, not all possibilities are explored equally often.
While the history of successful initial moves has been classified in opening theory \cite{Chess01}, about the mechanisms underlying the formation of fashionable openings not much is known \cite{Levene07}. 
%The choice of opening moves however relies more on popularities and opening preparations, rather than on calculation of winning chances
With the recent appearance of extensive databases 
playing habits have become accessible to quantitative analysis, making chess an ideal platform for analyzing human decision processes.

The set of all possible games can be represented by a directed graph whose nodes are game situations
and whose edges correspond to legal moves from each position (Fig.~1).
Every opening is represented by its move sequence as a directed path starting from the initial node.
We will differentiate between two game situations if they are reached by different move sequences. This way
the graph becomes a game tree, and each node $\sigma$ is uniquely assoiciated with an opening sequence.

Using a chess-database \cite{Database} we can measure the popularity $n_\sigma$ or weight of every opening sequence as the number of occurences in the database.
We find that the weighted game-tree of chess is self-similar and the frequencies $S(n)$ of weights follow a Zipf-Law \cite{Zipf49} 
\beq	\label{Eq:ZipfLaw}
	S(n) \sim n^{-\alpha}
\eeq
with universal exponent $\alpha=2$. 
Note, the precise scaling in the histogram of weight frequencies $S(n)$ and in the cumulative distribution $C(n)$ over the entire observable range (Fig.~2A).
Similar power-law distributions  with universal exponent have been identified in a large number of natural, economic and social systems
\cite{Sornette03,Mitzenmacher03,Newman05,Simon55,Pareto96,Willis22,Zipf49,Cox95,Klemm05} 
- a fact which has come to be known as Zipf- or Pareto law \cite{Pareto96,Zipf49}. 
If we count only the frequencies $S_d(n)$ of opening weights $n_\sigma$ after the first $d$ moves we still find broad distributions consistent with power-law behavior 
$S_d(n)\sim n^{-\alpha_d}$ 
(Fig.~2B). The exponents $\alpha_d$ are not universal, however, but increase linearly with $d$ (Fig.~2B, inset). 
The results are robust: similar power-laws could be observed in 
different databases and other board games, regardless of the considered game depth, constraints on player levels or the decade when the games were played.
Stretching over six orders of magnitude, the here reported distributions are one of the most precise examples for power-laws known today in social data sets.

\begin{figure}[t]
\includegraphics[height=2.95cm,clip=true]{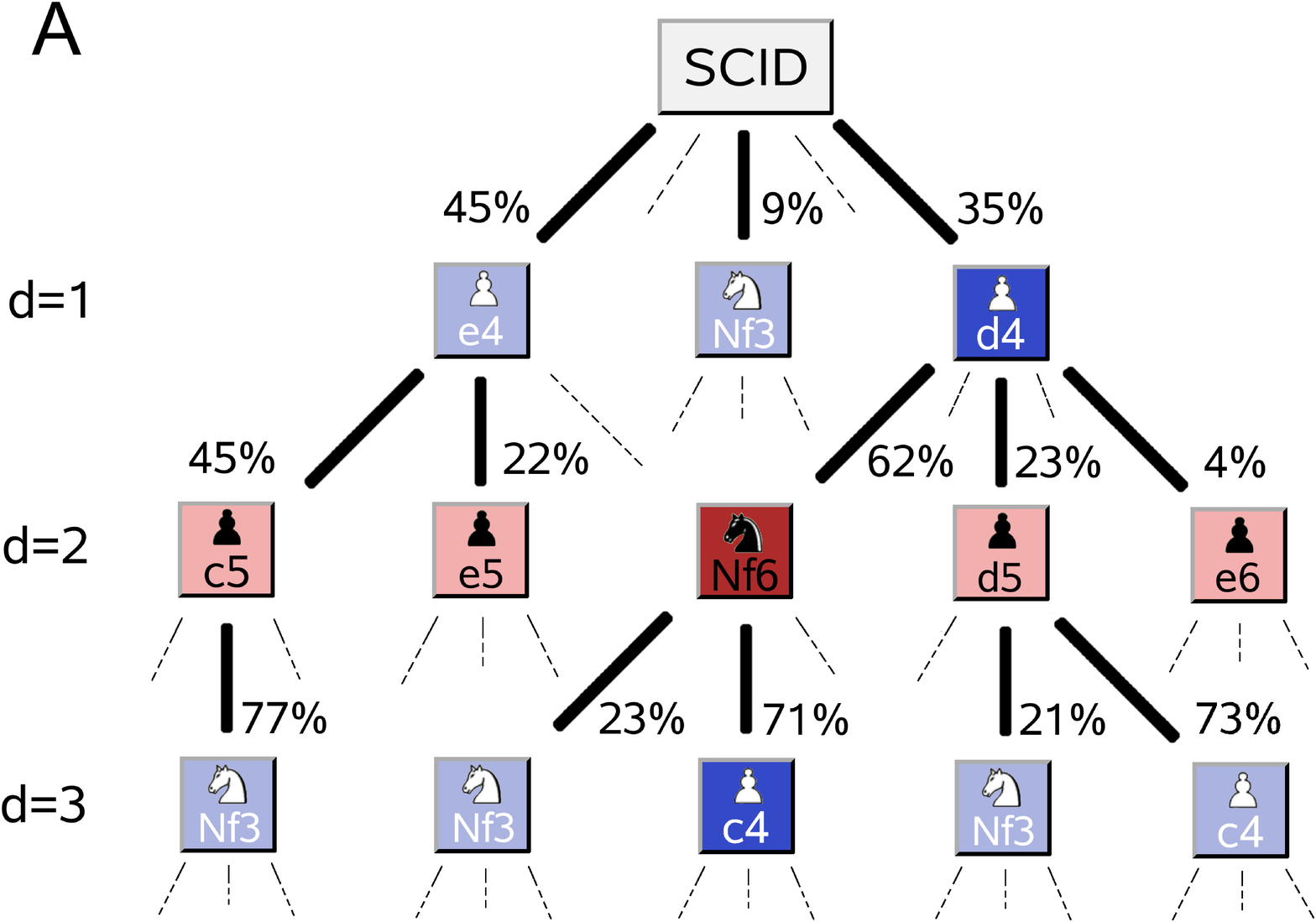} \hspace*{-4mm} 
\includegraphics[height=2.95cm,clip=true]{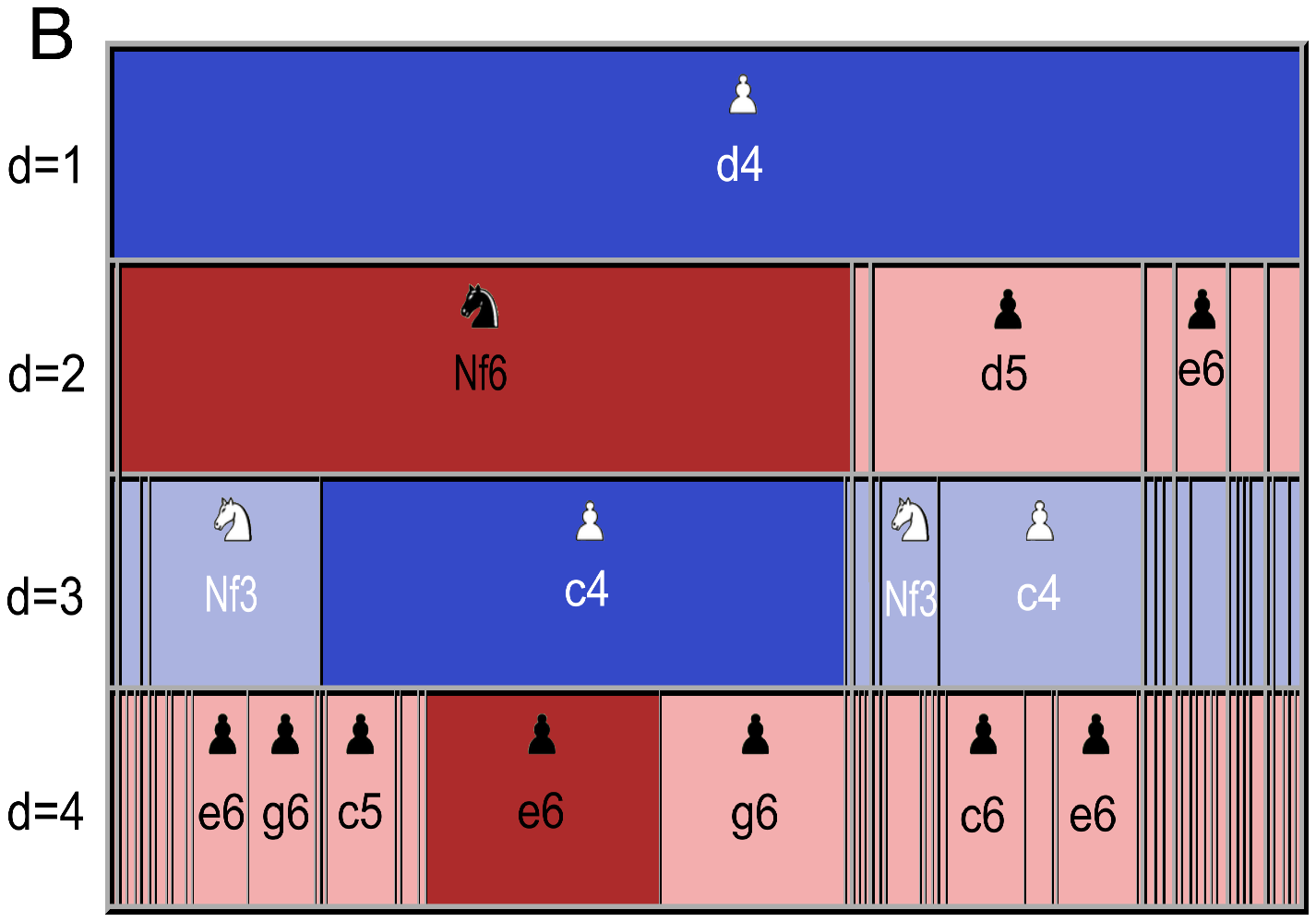}
\caption{%
{\bf A}) Schematic representation of the weighted game tree of chess based on the ScidBase \cite{Database} for the first three
half moves. Each node indicates a state of the game. Possible game
continuations are shown as solid lines 
together with the branching ratios $r_d$. Dotted lines symbolize other game
continuations, which are not shown. 
{\bf B}) Alternative representation emphasizing the
successive segmentation of the set of games,  here indicated for games following a
1.d4 opening until the fourth half move $d=4$. Each node $\sigma$ is represented by a box of a
size proportional to its frequency $n_\sigma$.
In the subsequent half move these games split into subsets (indicated vertically below)  according to the possible game continuations.
Highlighted in ({\bf A}) and ({\bf B}) is a popular opening sequence 1.d4 Nf6 2.c4 e6
(Indian Defense).}  \label{Fig:Fig1}
\end{figure}

As seen in (Fig.\ref{Fig:Fig1}) for each node $\sigma$ the weights of its subtrees define a partition of the integers $(1\dots n_\sigma)$. 
The assumption of self-similarity implies a statistical equivalence of the branching in the nodes of the tree. 
We can thus define the branching ratio distribution over the real interval $r\in[0,1]$ by the probability $Q(r|n)$ that a random pick from the numbers $1\dots n$ is in a subset of size smaller or equal to $rn$.
Taking $n$ to infinity $Q(r|n)$ may have a continuous limit $Q(r)$ for which we find the probability density function (pdf) $q(r)=Q'(r)$. 
If the limit distribution $q(r)$ of branching ratios exists it carries the fingerprint of the generating process.
For instance, the continuum limit of the branching ratio distribution for a Yule-Simon preferential growth process \cite{Simon55} in each node of the tree would be $q(r)\sim r^{\beta}$, where $\beta<0$ is a model specific parameter.
On the other hand, in a $k$-ary tree where each game continuation has a uniformly distributed random a-priori probability the continuum limit corresponds to a random stick breaking process in each node, yielding
 $q(r)\sim (1-r)^{k-2}$. 
For the weighted game tree of chess $q(r)$ can directly be measured from the database (Fig.~\ref{Fig:Fig3}A).  We find that $q(r)$ is remarkably constant over most of the interval but diverges with exponent $0.5$ as $r\to 1$, and is very well fitted by the parameterless arcsine-distribution  
\beq
q(r) = \frac{2}{\pi \sqrt{1-r^2}} \, .
%	q(r) = \frac{2}{\pi}\cdot\frac{1}{\sqrt{1-r^2}} \, .
\label{EqArcSine}
\eeq
The form of the branching ratio distribution suggests that in the case of chess there is no preferential growth process involved, but something entirely different which must be rooted in the 
decision process during the opening moves of a chess game \cite{Levene07}.

In the following we show that the asymptotic Zipf-Law in the weight frequencies arises independently from the specific form of the distribution $q(r)$, and hence, the microscopic rules of the underlying branching process.
Consider $N$ realizations of a general self-similar random segmentation process of $N$ integers, with paths $(\sigma_0,\sigma_1,\dots)$ in the corresponding weighted tree.  
In the context of chess each realization of this process corresponds to a random game from the database of $N$ games (e.g., dark shading  in Fig.~1). 
The weights $n_d=n_{\sigma_d}$ describe a multiplicative random process 
\begin{equation} 
	n_d=N\prod_{i=1}^{d}r_i  \quad,\quad n_0 = N
	\label{EqMulp}		
\end{equation}
where the branching ratios $r_d=n_d / n_{d-1}$ for sufficiently large $n_d$ are distributed according to $q(r)$ independent of $d$.
For lower values of $n_d$ the continuous branching ratio distribution is no longer a valid approximation and a node of weight one has at most one subtree, i.e. the state $n_d=1$ is absorbing.

\begin{figure}
\includegraphics[width=7.9cm,clip=true]{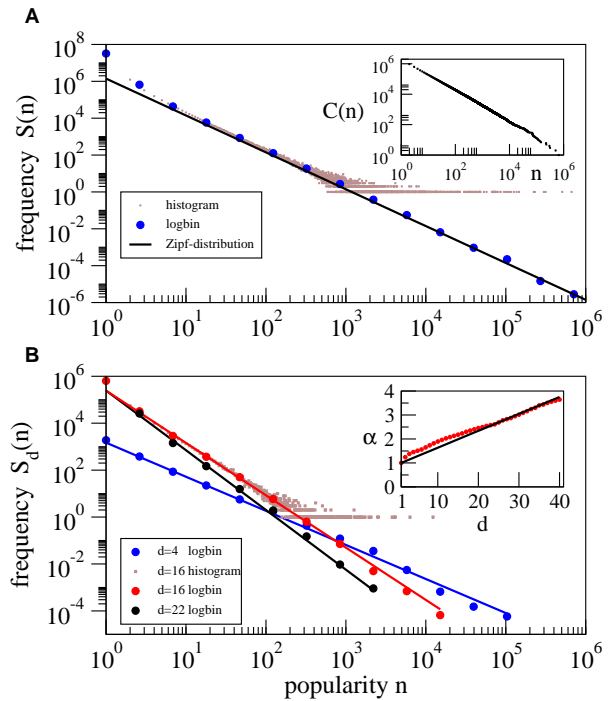}
\caption{{\bf A}) 
Histogram of weight frequencies
$S(n)$ of openings up to $d=40$ in the Scid database (brown dots) and with logarithmic binning (blue). 
A straight line fit (not shown) yields an exponent of $\alpha=2.05$ with a goodness of fit $R^2 > 0.9992$. For comparison, the Zipf-distribution Eq.~(\ref{EqUnivScLaw}) with $\mu=1$ is indicated as a solid line. 
Inset: number $C(n) = \sum_{m=n+1}^{N} S(m)$ of openings with a popularity $m>n$.
$C(n)$ follows a power-law with exponent $\alpha=1.04$ ($R^2 = 0.994$). 
{\bf B}) 
Number $S_d(n)$ of openings of depth $d$ with a given popularity $n$ for $d=16$ (brown dots) and histograms with logarithmic binning for  $d=4$ (blue), $d=16$ (red) and $d=22$ (black).  Solid lines are regression lines to the logarithmically binned data 
($R^2 > 0.99$ for $d<35$).  
Inset: slope $\alpha_d$ of the regression line as a function of  $d$ (red dots) and the analytical estimation 
Eq.~(\ref{EqAlphaD}) using $N=1.4 \cdot 10^6$  and $\beta=0$ (black solid line).} 
\label{Fig:Fig2}
\end{figure}

To calculate the probability density function (pdf) $p_d(n)$ of the random variable $n_d$ after $d$ steps 
it is convenient to consider the log-transformed variables $\nu = \log(N/n)$ and $\rho = -\log r$. 
The corresponding process $\{\nu_d\}$ is a random walk 
$\nu_d = \sum_{i=1}^d \rho_i $  with non-negative increments $\rho_i$
and its pdf $\pi_d(\nu)$ transforms as  $n\, p_d(n) = \pi_d(\nu)$. 
An analytic solution can be obtained for the class
\beq 
\label{EqGenPowBeta}
q(r)=(1+\beta) \, r^\beta,  \qquad 0\leq r \leq1 \, ,
\eeq
of power-law distributions, which typically arise in preferential attachment schemes.
In this case the jump process $\nu_d$ is Poissonian and distributed according to a gamma distribution
$ \pi_d(\nu) = \tfrac{(1+\beta)^d}{(d-1)!}\nu^{d-1} e^{-(1+\beta)\nu}$.
After retransformation to the original variables and noting that
from the probability $p_d(n)$ for a single node at distance $d$ to the root to have the weight $n$ 
one obtains the expected number $S_d(n)$ of these nodes in $N$ realizations of the random process as $ S_d(n)= N
p_d(n) / n$, and in particular
\beq \label{EqLogGamma}
S_d(n) = \frac{(1+\beta)^d}{N(d-1)!}\left(\log\frac{N}{n}\right)^{d-1}\left(\frac{N}{n}\right)^{1-\beta}
\, .
\eeq
The functions $S_d(n)$ are strongly skewed and can exhibit power-law like scaling over several decades.  A logarithmic expansion  for $1< n \ll N$ shows that they approximately follow a scaling law $S_d(n) \sim n^{-\alpha_d}$ with exponent
\beq	\label{EqAlphaD}
	\alpha_d = (1-\beta) + \frac{1}{\log N} \,(d-1) \,.
\eeq
The exponent $\alpha_d$ is linearly increasing with the game depth $d$ and with a logarithmic finite size correction which is in excellent agreement with the chess database (Fig.~2B, inset). 
Power-laws in the stationary distribution of random segmentation and multiplicative processes have been reported before \cite{Sornette03} and can be obtained by introducing slight modifications, such as
reflecting boundaries, frozen segments, merging or reset events \cite{Sornette98,Krapivsky00,Banavar04}. 
In contrast, the approximate scaling of $S_d(n)$ in Eq.~(\ref{EqLogGamma}) is fundamentally different, as our process does not admit a stationary distribution. The exponents $\alpha_d$ increase due to the finite size of the database.

As shown in Fig.~3B we find excellent agreement between the weight frequencies $S_d(n)$ in the chess database and direct simulations of the multiplicative process, Eq.~(\ref{EqMulp}) using the arcsine distribution Eq.~(\ref{EqArcSine}). 
If the branching ratios are approximated by a uniform distribution $q(r)=1$
the predicted values of $S_d(n)$ are systematically too small, since a uniform distribution yields a larger flow into the absorbing state $n^*=1$ than observed in the database. Still, due to the asymtotic behavior of $q(r)$ for $r\to 0$, this approximation
yields the correct slope in the log-log plot so that the exponent $\alpha_d$ can be
estimated quite well based on  Eq.~(\ref{EqAlphaD}) with $\beta=0$.

\begin{figure}[t]
\includegraphics[width=7.5cm,clip=true]{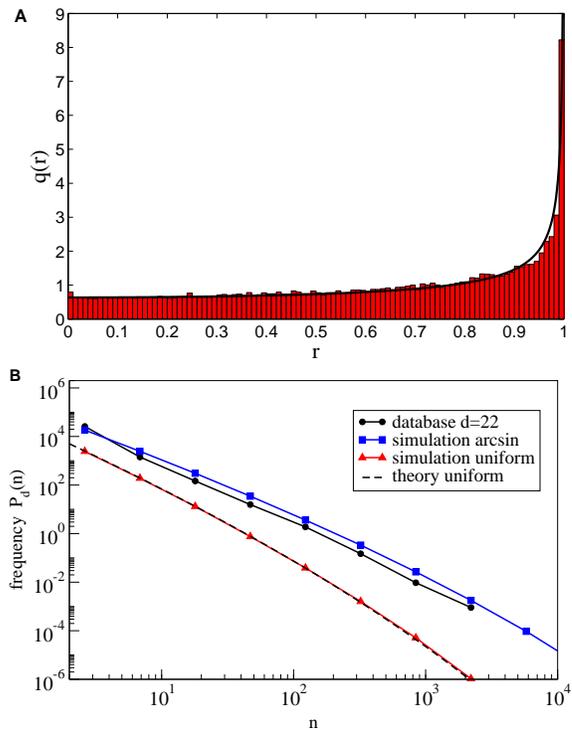} 
\centerline{\includegraphics[width=7.5cm,clip=true]{./fig3b.eps}}
\caption{
{\bf A}) Probability density $q(r)$ of branching
ratios $r$ sampled from all games in the Scid database with a bin size of $\Delta r
=0.01$ (red bars)  and arcsine distribution  Eq.~(\ref{EqArcSine}) (black solid line). 
Every edge of the weighted game tree, from nodes of size $n_{d-1}$
to $n_d$, contributes to the bin corresponding to $r=n_d/n_{d-1}$ with weight $r$. 
We disregarded clusters with $n_d<100$ so that, in principle, a cluster could contribute to any of the bins.
We found $q(r)$ to be depth-independent.
{\bf B}) Distribution of opening popularities $S_d(n)$ for $d=22$ obtained from the Scid database (black) and from a direct simulation of the multiplicative process Eq.~(2),  whith branching ratios $q(r)$ taken from a uniform (red) or arcsine (blue)  distribution.
Further indicated is the theoretical result Eq.~(\ref{EqLogGamma}) (dashed line).
Similar results are obtained for other values of $d$.
} \label{Fig:Fig3}
\end{figure}

By observing that $S_d(n)$ in Eq.~(\ref{EqLogGamma}) is the \mbox{$d$-th} term in a series expansion of an exponential function, we find the weight distribution in the whole game tree as $S(n)=\sum_d S_d(n)$ to be an exact Zipf-Law.
For branching ratio distributions $q(r)$ different from Eq.~(\ref{EqGenPowBeta}) the weight frequencies 
are difficult to obtain analytically.
But using renewal theory \cite{Feller71} the scaling 
can be shown to hold asymptotically for $n\ll N$ and a large class of distributions $q(r)$.
For this, note that the random variable $\tau(\nu)=\max(d:\nu_d<\nu)$ is a renewal process in $\nu$. 
The expectation $\mathbf{E}[\tau(\nu)]$ is the corresponding renewal function
related to the distributions of the $\nu_d$ as
$	\sum_{d=1}^{\infty} \mathrm{Prob}(\nu_d<\nu) = \mathbf{E}[\tau(\nu)] $.
If the expected value $\mu=\mathbf{E}[\rho]=\mathbf{E}[-\log r]$ is finite and
positive (e.g., for the distribution (\ref{EqGenPowBeta}) $\mu=1/(1+\beta)$),
the renewal theorem provides
\beq \label{EqReneqTh}
	\lim_{\nu\to\infty} \frac{d}{d\nu} \mathbf{E}\left[\tau(\nu)\right] = \frac{1}{\mu} \, .
\eeq
Thus, we obtain
$\lim_{\nu\to\infty} \sum_{d=1}^{\infty} \pi_d(\nu) = \frac{1}{\mu} $
and finally
\beq \label{EqUnivScLaw}
\lim_{\frac{n}{N} \to 0}	S(n)  = \frac{N}{\mu n^2} \, .
\eeq
Thus, the multiplicative random process (Eq.~\ref{EqMulp}) with any well behaving branching ratio distribution $q(r)$ on the intervall $[0,1]$ always leads to an asymptotically universal scaling for $n\ll N$ (compare also the excellent fit of
Eq.~(\ref{EqUnivScLaw}) to the chess data in Fig.~2a).
In \cite{Klemm05} the same Zipf-Law scaling was found for the sizes of the directory trees in a computer cluster. The authors propose a growing mechanism based on linear preferential attachment. Here we have shown that the exponent $\alpha=2$ for the weight distribution of subtrees in a self-similar tree is truly universal in the sense that it is the same for a much larger class of generating processes and not restricted to preferential attachment or growing.

\begin{figure}
\includegraphics[width=8.7cm,clip=true]{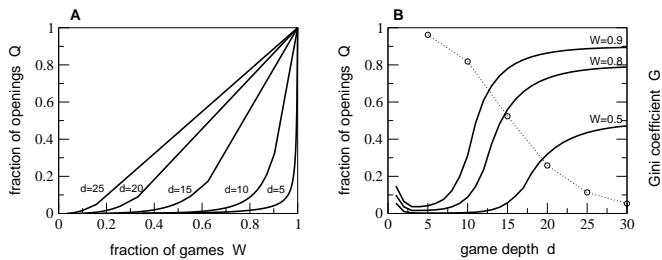}
\caption{Inequality \cite{Gastwirth72,Newman05}
of the distribution $S_d(n)$.  
{\bf A}) 
Proportion $W$ of games that is concentrating in the fraction $Q$ of the most popular openings, for several levels of the game depth $d$.
{\bf B})
$Q$ as a function of $d$ for three different values of $W$ (solid lines)
and Gini-coefficient $G=1-2\int_0^1 Q(W)\, dW$ as a function of game depth (dotted line).
} \label{Fig:Fig4}
\end{figure}

There are direct implications of our theory to general composite decision processes, where each action is assembled from a sequence of $d$ mutually exclusive choices.  
What in chess corresponds to an opening sequence, may be a multivariate strategy or a customized ordering in other situations. The question how such strategies are distributed is important for management and marketing \cite{LongTail}.
One consequence of our theory is, that in a process of $d$ composite decisions the distribution $S_d(n)\sim n^{-\alpha_d}$ of decision sequences, or strategies, which occur $n$ times shows a transition from low exponents $\alpha_d\le 2$, where a few strategies are very common, to higher exponents $\alpha_d> 2$, where individual stategies are dominating. This is due to the divergence of the first moment in power-laws with exponents smaller than two \cite{Newman05}. 
From (Eq.~\ref{EqAlphaD}) the critical number $d_{cr}$ of descisions at which this transition occurs depends logarithmically on the sample size $N$ and on the leading order $\beta$ of $q(r)$ near zero as
\beq
	d_{cr}=1+(1+\beta)\log N	\, .
	\label{EqCriticDepth}
\eeq

Applied to the chess database with $N=1.4 \cdot 10^6$ we obtain $d_{cr}\approx15$ (see also Fig.~4 and Fig.~2B inset).
This separates the database into two very different regimes: in their initial phase $(d<d_{cr})$ the majority of chess games is distributed among a small number of fashionable openings
(for $d=12$, for example, 80\% of all games in the database are concentrated in about 23\% of the most popular openings), whereas beyond the critical game depth rarely used move sequences are dominating such that in aggregate they comprise the majority of all games (Fig.~4).
Note, that this result arises from the statistics of iterated decisions and does not indicate a crossover of playing behavior with increasing game depth.

Our study suggests the analysis of board games as a promising new perspective for statistical physics.
The enormous amount of information contained in game databases, with its evolution resolved in time and in relation to an evolving network of players, provide a rich environment 
to study the formation of fashions and collective behavior in social systems.

We are indebted to Andriy Bandrivskyy for invaluable help with the data analysis.

%%%%%%%%%%%%%%%%%%%%%%%%%%%%%%%%%%%%%%%%%%%%%%%%%%%%%%%%%%%%%%%%%%%%%%%%

\end{document}